\documentclass[a4paper,11pt]{article}

\usepackage{jheppub}
\usepackage{amsmath}
\usepackage{caption}
\usepackage{subcaption}
\usepackage{amsfonts}
\usepackage{graphicx}
\usepackage{epstopdf}
\usepackage{bm}
\usepackage{slashed}
\usepackage{ulem}

\usepackage{hyperref}
\hypersetup{
    colorlinks=true,
    linkcolor=blue,
    filecolor=magenta,      
    urlcolor=cyan,
    pdftitle={Overleaf Example},
    pdfpagemode=FullScreen,
    }

\newcommand{\be}[0]{\begin{equation}}
\newcommand{\ee}[0]{\end{equation}}

\newcommand{\tr}{\textrm{tr }}

\newcommand{\R}{{\mathbb{R}}}

\newcommand{\s}{{S}}

\newcommand{\dd}{{\text{d}}}

\newcommand{\order}[1]{\mathcal{O}\left( #1 \right)}
\newcommand{\const}{\text{const}}
\newcommand{\pd}{\partial}
\newcommand{\Tr}{\textrm{Tr }}

\newcommand{\diag}{\textrm{diag }}
\newcommand{\critical}[1]{#1^{\text{cr}}}

\begin{document} 

\title{On Emergent Directions in Weakly Coupled, Large N$_c$ $\mathcal{N}=1$ SYM}

\author[a]{Baiyang Zhang,}
\emailAdd{byzhang@henu.edu.cn}

\author[b]{and Aditya Dhumuntarao}
\emailAdd{dhumu002@umn.edu}

\affiliation[a]{Institute of Contemporary Mathematics, School of Mathematics and Statistics, Henan University, Kaifeng, Henan 475004, P. R. China}

\affiliation[b]{Quantum Information Science, Sandia National Laboratories, Livermore, CA 94550}
\date{\today}


\abstract{

The $SU(N)$ Yang-Mills theory compactified on $\R^3 \times S^1_L$ with small $L$ has many merits, for example the long range effective theory is weakly coupled and adopts rich topological structures, making it semi-classically solvable. 
Due to the $SU(N) \to U(1)^{N-1}$ symmetry breaking by gauge holonomy, the low-energy effective theory can be described in terms of unbroken $U(1)$ photons and gauge holonomy.
With the addition of $N_f$ adjoint light fermions, the center symmetry breaking phase transition can be studied using the twisted partition function, i.e., fermions with periodic boundary conditions, which preserve the supersymmetry in the massless case.
In this paper, we show that in the large-$N$ abelian limit with $N_f=1$ and an $N$-independent W-boson mass, the long-range $3$d effective theory can be regarded as a bosonic field theory in $4$d with an emergent spatial dimension. The emergent dimension is flat in the confining phase, but conformally flat in the center-symmetry broken phase with a $\mathbb{Z}_2$ reflection symmetry. The center symmetry breaking phase transition itself is due to the competition between instanton-monopoles, magnetic and neutral bions controlled by the fermion mass, whose critical value at the transition point is given analytically in the large $N$ limit.\\
}


\maketitle
\flushbottom

\section{Introduction}
~

Yang-Mills theories in four spacetime dimensions simplify in the limit that the number of colors $N$ becomes large~\cite{tHooft:1973}.  While typically they do not simplify enough to become solvable, by now a number of general features of the large $N$ limit has been understood. Since the 1980s, volume independence or the (twisted) reduction mechanism at large $N$ in decompactified lattice gauge theory have been extensively studied~\cite{Eguchi1982,Bhanot1982,Neuberger2020,GonzalezArroyo1983,UnsalYaffe2010}. It is now well understood that in a $d$-dimensional lattice gauge theory, as long as the $U(1)^{d}$ center symmetry remains unbroken, the standard Wilson lattice theory reduces to the Eguchi-Kawai (EK) model, defined on a single hypercube with specific boundary conditions, and the loop equations become volume-independent. Moreover, the effective action can be expressed in terms of the eigenvalues of the link matrices, which resemble lattice momenta. This resemblance underpins the key concept of color-momentum transmutation and the quenched momentum prescription, which apply to both lattice and continuum theories, see~\cite{Gross1982,ArroyoOkawa2014} and references therein. Also, see~\cite{Arkani-Hamed:2001kyx} for a similar resemblance between group indices and spatial lattice. 

Another startling feature is that in many cases (and conjecturally, in all cases) the 't Hooft large $N$ limit\footnote{The 't Hooft large $N$ limit is defined by taking $N$ to infinity while fixing the number of quark flavors, the box size, the bare quark masses, and the 't Hooft coupling. } of a quantum field theory can be understood in terms of a string theory living in a higher-dimensional spacetime.  The field-theoretic and gravitational descriptions are related by a strong coupling/weak coupling duality, so that the large $N$ QFT is strongly coupled (the 't Hooft coupling $\lambda = g^2 N$ is large) when the string theory is weakly coupled ($g_s = 0$) and weakly curved, meaning that the curvature radius is large in string units~\cite{Maldacena:1997re,Gubser:1998bc,Witten:1998qj}. 

One might wonder whether there are some $4$d gauge theories with weakly-coupled, solvable large $N$ limits, and if so, whether there is any sense in which the physics of such large $N$ limits can be usefully described by some model which lives in a different spacetime than the original one. It was observed that, along with some other examples, 4d $\mathcal{N}=1$ super-Yang-Mills theory with gauge group $SU(N)$ has a large $N$ limit with this property, see Ref.~\cite{DOUGLAS1995271,Unsal2009,Cherman:2016jtu}. Before commenting on its solvable large $N$ limit, we should say that  this theory has a standard 't Hooft large $N$ {limit where one takes $N\to \infty$ with the strong scale $\Lambda$ and all other parameters held fixed}. While it is reasonable to expect that this large $N$ limit has some sort of holographic string-theoretic description, it is expected to be quite complicated, involving either string theory in highly curved spacetimes, or a simpler supergravity theory at the cost of describing a field theory with $\mathcal{N}=1$ supersymmetry but with light adjoint matter multiplets~\cite{Aharony:1999ti,Polchinski:1992vg,Girardello:1999bd,Petrini:2018pjk}.

But there is another interesting way to take the large $N$ limit.  Suppose that we place SYM theory on the manifold $\mathbb{R}^3 \times S^1$, where $S^1$ has a circumference $L$, and we assume that the fermions have periodic boundary conditions on $S^1$. The abelian large $N$ limit on $\mathbb{R}^3 \times S^1$ was first introduced in Ref~\cite{UnsalYaffe2008}. However, the large $N$ limit with $\Lambda$ held fixed is no longer
unique:  the theory now has the extra parameter $\eta \equiv NL\Lambda$.  If we hold $L\Lambda$ fixed as $N\to \infty$, we get the standard 't Hooft large $N$ limit.  But we can also decide to hold $\eta$ fixed as $N\to \infty$, so that $L\Lambda \sim 1/N$.   

If $\eta$ is small, it is known that the theory is weakly coupled at all length scales even at large $N$~\cite{Poppitz:2012sw,Anber:2014sda,Poppitz:2012nz,Anber:2015wha}.  The reason this happens is that when $\eta \ll 1$, the $SU(N)$ gauge group is Higgsed to $U(1)^{N-1}$, and the $W$ boson masses are
set by the scale $\Lambda/\eta \gg \Lambda$.  This phenomenon is tied up with the fact that $\mathcal{N}=1$ $SU(N)$ SYM has a $\mathbb{Z}_N$ 1-form ``center'' symmetry, and this symmetry is believed not to be spontaneously broken for any $L$ in the setup described above.  When $L \Lambda \ll 1$, this can be seen from an analytic calculation.  The expectation values of Polyakov loops around $S^1$
vanish, $\langle \Tr \mathcal{P}e^{i n\int_{S^1} A} \rangle = 0$ for $n \neq 0
\textrm{ mod } N$, and the gauge field in the $S^1$ direction, say $A_4$, acts like an adjoint Higgs field. The fact that $m_W \gg \Lambda$ means the gauge coupling stops running at a high scale $\mu \sim m_W\gg \Lambda$, and implies that theory will be weakly coupled
at all scales provided $\eta \ll 1$.

An important feature of this construction is that at finite $N$, the small-circle and large-circle theories have precisely the same realization of all global symmetries~\cite{Davies:2000nw,Seiberg:1996nz}.  As already mentioned above, it is believed that the $\mathbb{Z}_N$ 1-form ``center'' symmetry is not spontaneously broken either at large or small $L$, and the $\mathbb{Z}_{2N}$ chiral symmetry is spontaneously broken for both small and large $L$. This makes it plausible the large and small $L$ (that is, large and small $\eta$) theories are smoothly connected to each other.  Indeed, this conjecture is supported by lattice calculations, see e.g.~\cite{Bergner:2018unx,Athenodorou:2020clr}.  These ideas can be extended to quantum theories without any supersymmetry, and gives a window on studying non-perturbative dynamics in a wide variety of QCD-like theories.  In all of these cases, the idea is that by studying the tractable theories with small $\eta$ we can hope to get insight into the strongly-coupled physically-interesting version of theory with large
$\eta$.  This is the same motivation adopted in the huge number of studies which consider highly supersymmetric cousins of interesting non-supersymmetric theories, but without the necessity to assume supersymmetry at the very start.

However, in this paper we will focus on $\mathcal{N}=1$ $SU(N)$ SYM as a first step, although our analysis use techniques which can also be applied to non-SUSY QFTs.  It would be natural to guess that the small-$\eta$ limit of theory on $\mathbb{R}^3 \times S^1$ can be described by a $3$d effective field theory, since it is formulated on a small circle.  But at large $N$, it turns out that the physics cannot be described by a $3$d EFT. Instead, one can use a $4d$ EFT. This EFT is formulated on a spacetime of the form $\mathbb{R}^3 \times \tilde{S}^{1}$, where $\tilde{S}^1$ is an emergent direction of size $\tilde{L}$.  The EFT contains a gapless mode with $z=2$ Lifshitz scale invariance, and its bosonic sector can be represented by the action
\begin{align}
  S_b = \int_{\R^3} d^3{x} \int_0^{\tilde{L}} dy\, \left( |\partial_{\mu} \Phi|^2 + |\partial_{y}^2 \Phi|^2\right)
  \label{eq:lifshitz_action}
\end{align}
where $\mu = 0,1,2,3$, $\Phi$ is a complex scalar field and $\tilde{L}$ is the size of the emergent dimension.  Note that the emergent direction adopts translation symmetry, and the spacetime $\mathbb{R}^3 \times \tilde{S}^{1}$ is as flat as the original physical spacetime $\mathbb{R}^3 \times S^{1}$. As explained in Ref.~\cite{Cherman:2016jtu}, after
taking into certain non-perturbative effects, the $3$d $U(1)^{N-1}$ effective field theory that describes the physics at small $\eta$ can be interpreted as a theory with a discretized extra dimension on a lattice with $N$ sites, and the degrees of freedom moving on this lattice can be related to the photons of
$U(1)^{N-1}$ and their superpartners. The translation symmetry on the emergent lattice is generated by the $0$-form part of center symmetry, and the fact that the center symmetry is \emph{not} spontaneously broken is in one-to-one correspondence with the fact that translation symmetry in the emergent dimension is not spontaneously broken.

Here we consider what happens to the picture above if we turn on a mass $m$ for
the adjoint quark in $SU(N)$ super-Yang-Mills theory.  This explicitly breaks supersymmetry and the $\mathbb{Z}_{2N}$ chiral symmetry, but it does not explicitly break the center symmetry.  However, when the quark mass is large enough, the theory becomes equivalent to pure non-supersymmetric Yang-Mills
theory, and pure YM theory spontaneously breaks center symmetry on $\mathbb{R}^3 \times S^1_{L}$ for small enough $L$.  So as we increase $m$, we should expect that center symmetry breaks spontaneously at some critical value $\critical{m}$.  It is known that this quantum phase transition is first-order~\cite{Cherman:2016jtu, Poppitz:2012nz}.

Our goal here is to explore the large $N$ limit of $\mathcal{N}=1$ $SU(N)$ SYM perturbed by the small quark mass $m$. When $m = 0$, the low-energy description is given by Eq.~\eqref{eq:lifshitz_action}, and was already understood in Ref.~\cite{Cherman:2016jtu}.  We will show that when $0 < m < \critical{m}$ for certain transition fermion mass $\critical{m}$, the
low-energy effective action continues to be a scale-invariant theory with a flat emergent spatial direction, but with $z =1$ rather than $z=2$.  Once $m > \critical{m}$, however, we will see that the emergent $4d$ spacetime becomes \emph{warped} {and a gap emerges in the emergent spacial $S^1$ as a result of the spontaneous breaking of $\mathbb{Z}_N$ (discrete-)rotational symmetry.}  
The low-energy effective action takes the form 
\begin{align}
  S = \int_{\mathbb{R}^3} d^3 x\,   \int_{\mathbb{I}}d y \sqrt{|{g}|} g^{ab}\partial_a\phi \partial_b\phi
  \label{eq:curved_action}
\end{align}
where $\mathbb{I}$ is an open interval on $\mathbb{R}$ and
\begin{align}
  \dd s^2 = g_{ab}\dd x^a \dd x^b = 
  f(y)\delta_{ij}\dd x^i\dd x^j + f(y)^{-1}\dd y^2,\hspace{.25in} f^2(y) 
  = \frac{\tilde{c}}{8}\left[1-4\left(\frac{y}{\tilde{L}}-\frac{1}{2}\right)^2\right] 
\end{align}
where $\tilde{c}$ is proportional to $m$.  We will then discuss the properties of this rather peculiar metric, and comment on some implications for the physics of
the gauge theory.


\section{Mass-deformed $\mathcal{N}=1$ SYM theory}
The model under consideration is described by the action
\begin{align}
  S = \int d^4x\, \left[ \frac{1}{4g^2} \tr f_{\mu\nu} f^{\mu \nu} + \bar{\psi}\left(i\slashed{D} - m \right)\psi \right]\, ,
\end{align}
where $f_{\mu\nu}$ is the $SU(N)$ field strength, $\psi$ denotes an adjoint representation Majorana fermion. The $SU(N)$ generators are normalized such that $\tr t_a t_b = \delta_{ab}$.  We work in Euclidean signature with $\theta=0$, the spacetime manifold is defined as $\mathbb{R}^3 \times S^1_L$, where $L$ is the circumference of $S^1$. The model exhibits a $\mathbb{Z}_N$ 1-form symmetry acting on Wilson loops according to $W_R[C] = \text{tr}_R \left\{\mathcal{P} e^{i \int_C a}\right\}$, where $C$ is a closed curve.  For our application here, we recall that this $1$-form ``center'' symmetry acts on Polyakov loops by multiplying them by $\mathbb{Z}_N$ phases:
\begin{align}
  W_{R}(S^1) \to e^{2\pi i k/N}   W_{R}(S^1), \quad k = 0,\dots, N-1 . 
\end{align}
The value of $\langle W_{\rm F}(C)\rangle$ is an order parameter for the realization of center symmetry.  When $m=0$, it is known that this symmetry is not spontaneously broken, and $\langle W_{\rm R}(C)\rangle = 0$ for all representations with non-vanishing $N$-ality.  We will summarize the calculation that leads to this result below.  When the theory is weakly coupled, the eigenvalues of the Polyakov loop have small fluctuations, and if center symmetry is not spontaneously broken, when $L$ is sufficiently small one can implement the condition $\langle W_{\rm R}(S^1)\rangle = 0$ by writing $W_{R}(C) = \text{tr}_R \Omega$, where $\Omega = e^{i\zeta} \diag(1, \omega, \omega^2, \cdots, \omega^{N-1})$, $\omega = e^{2\pi i /N}$ and $\zeta$ is chosen such that $\det \Omega = 1$.

Physically, the eigenvalues of the Polyakov loop have small fluctuations and one can describe the long-distance physics using a 3d effective field theory.  The $S^1$ component of the gauge field enters this EFT as an adjoint Higgs field, and the center-symmetry-preserving expectation value for $\Omega$ serves to Higgs the gauge group from $SU(N)$ to $U(1)^{N-1}$.  The smallest W-boson mass is $m_W \sim \frac{2\pi}{NL}$, and the weak-coupling description is self-consistent provided that $m_W$ is much larger than the strong scale $\Lambda$, which can be written as ~\cite{Poppitz:2012nz,Poppitz:2021cxe}
\begin{align}
 \Lambda^3 = \mu^3 \frac{16\pi^2}{3Ng^2(\mu)}\exp\left( -\frac{8\pi^2}{g^2(\mu)N} \right).
\end{align}

To describe the long-distance physics, it is very helpful to use 3d Abelian duality to trade the $N-1$ ``photon'' fields $A^{i}_{\mu} $ for $N-1$ dual scalar fields $\sigma_i, i = 1, \cdots N-1$~\cite{Turner:2019wnh}.   The dual photons are massless to all orders in perturbation theory, but they pick up a small mass (exponentially small compared to $m_W$) due to the proliferation of certain non-BPS magnetically-charged finite-action field configurations called magnetic bions~\cite{Unsal2009,Poppitz:2022}.  The supersymmetry of the model at $m=0$ implies that there is no potential for the holonomy to any order in perturbation theory, which means that the $N-1$ fields representing holonomy eigenvalue fluctuations are also light.  Finally, the model also has some light fermionic modes which are necessary for the low-lying spectrum to fit into an appropriate 3d $\mathcal{N}=2$ supermultiplet when $m = 0$.
 
We assemble  the dual photons and holonomy fluctuation fields into $N$-vectors $\bm{\sigma}$ and $\bm{\phi}$ (this simplifies the notation at the cost of adding two exactly decoupled fields), 
and then follow  Ref.~\cite{Poppitz:2012nz} and write
\begin{equation}
  \bm{\phi} = \frac{2\pi}{N}\bm{\rho} + \frac{g^2}{4\pi}\bm{b'}, \quad
  \bm{\sigma} = \frac{2\pi k }{N}\bm{\rho}  + \bm{\sigma '},
\end{equation}
where $k = 1,\cdots,N$ and $\bm\rho$ is the Weyl vector, with the supersymmetric ground states identified. {When center symmetry is not spontaneously broken}, the field $b'$ represents the fluctuation of the phases of the gauge holonomy and $\sigma'$ is the fluctuation of the dual photon field.  The integer $k$ labels the $N$ vacua arising from the spontaneous breaking of the $\mathbb{Z}_N$ chiral symmetry of the model when $m=0$.  We will be interested in the behavior of the model when $m >0$, so we will set $k=0$ from here onward. 
Note that $\bm{b}',\bm{\sigma}'$ are periodic
\be
  \bm{b}' \sim \bm{b}'+\frac{8\pi^2}{g^2} \bm{\omega},\quad
  \bm{\sigma}' \sim \bm{\sigma}'+2\pi \bm{\omega}.
\ee

We assume that the Cartan generators of SU(N) are normalized as $\tr{H_i H_j} = \delta_{ij}$. The root vectors are $\bm{\alpha_i} = (0,\cdots, 1, -1, 0 ,\cdots, 0)$ where $1$ is the $i$-th component.  The affine root $\alpha_N = - \sum_{i=1}^{N-1} \alpha_i$, $\bm{\omega}$ and $\bm{\omega}^*$ are the fundamental weights and co-weights respectively and the Weyl vector is given by $\bm{\rho} = \sum_i \bm{\omega}_i$. The fundamental weights are normalized as $\bm{\omega}_i\cdot \bm{\alpha}_j = \delta_{ij}$, thus $\bm{\rho}\cdot \bm{\alpha}_i = 1$ for $i = 1, \cdots, N-1$ and $\bm{\rho}\cdot \bm{\alpha}_N = 1-N$. 
Finally, we note that the gauge holonomy in terms of $\bm{b}'$ reads
\be
  \Omega = \exp \left( i\frac{2\pi}{N}\bm{H}\cdot\bm{\rho}
  + i \frac{g^2}{4\pi} \bm{H}\cdot\bm{b}' \right).
  \label{eq:Omega}
\ee

The ``GPY'' effective action, incorporating both perturbative and nonperturbative contributions on a thermal circle, was first derived in the seminal work~\cite{Gross:1980br}. Transitioning from a thermal to a nonthermal circle requires a shift in perspective, the effective potential in the latter case was given in~\cite{Unsal2009}. The effective potential for $n_f$ flavors of adjoint fermions was subsequently given in~\cite{UnsalYaffe2010}. At one loop, this effective potential takes the form
\begin{align}
  V_{\rm eff}(\Omega) = \frac{2}{\pi^2L^3} \sum_{n=1}^{\infty}\left[-1 + \tfrac{1}{2} (n L m)^2 K_2(n L m)\right] \frac{|\tr \Omega^n|^2 }{n^4}\,,
\end{align}
where we have normalized it as appropriate within a 3d EFT, and $K_2$ is the modified Bessel function of the second kind.  When $m=0$, the model we are studying has $\mathcal{N}=1$ supersymmetry, and the contributions of bosons and fermions to the effective potential cancel to all orders in perturbation theory.  When $m$ is small but non-zero, the one-loop effective potential becomes 
\begin{align}
  V_{\rm eff}(\Omega) = -V_\text{p} \sum_{n=1}^{\infty} \frac{|\tr \Omega^n|^2 }{n^2} + \mathcal{O}(m^3)\,,
\end{align}
where 
\begin{align}
  V_{\text{p}} = \frac{m^2}{2\pi^2L} = \frac{\Lambda  m^2 \text{N}}{2 \pi ^2 \eta }.
\end{align}
Note that the leading perturbative contribution to the holonomy effective potential scales as $m^2$.  The crucial point is that non-perturbative effects generate contributions to the holonomy effective potential which are proportional to $m^0$ and $m^1$~\cite{Poppitz:2013}.  To describe these contributions, let us assume that the eigenvalues of the holonomy are not coincident; the subsequent
calculation will show that this assumption is self-consistent.  Then there are $N$ BPS monopole-instantons, and when center symmetry is not spontaneously broken they all have action $8\pi/(g^2N)$.  Each of the monopole-instantons carry two fermion zero modes when $m=0$, and so they cannot contribute to the
effective potential, which is bosonic.  However, they can form non-BPS `bound states'
that do not have any fermion zero modes, which are magnetic and neutral bions.
It is known that the proliferation of these topological defects gives a
contribution to the holonomy effective potential in $\mathcal{N}=1$ SYM
theory\cite{Poppitz:2021cxe,Poppitz:2012nz}.  
When the fermion mass is turned on, there are two basic effects.  First, the fermion zero modes are lifted, and monopole-instantons can contribute
some terms proportional to $m$ to the holonomy effective potential.  Second,
boson and fermion loops no longer cancel, so there is also a one-loop
contribution to the holonomy effective potential.  However, since this
perturbative contribution is proportional to $m^2$, it is
negligible compared to the monopole-instantons contribution. 

Dropping the primes in $\bm\phi',\bm\sigma'$, the non-perturbative effective potential for the gauge holonomy with fermion mass $m$ is known to take the form
\begin{align}
\notag
  V_{\text{np}} = & V_{\text{b}} \left[ \sum_{i=1}^N e^{-2\bm{\alpha_i} \cdot \bm{b}} -e^{-(\bm{\alpha_i}+\bm{\alpha_{i+1}}) \cdot \bm{b}}\cos [(\bm{\alpha_i}-\bm{\alpha_{i+1}})\cdot \bm{\sigma}]\right] \\
  & -V_{\text{m}} \left[ \sum_{i=1}^{N}\left( 1+\frac{g^2 N}{8\pi^2}\bm{\alpha_i} \cdot \bm{b} \right)
                e^{-\bm{\alpha_i} \cdot \bm{b}}
                \cos \left( \bm{\alpha_i} \cdot \bm{\sigma} + \frac{2\pi k + \theta}{N} \right) \right]
  \label{eq:nppot}
\end{align}
where the first term in the first line is the contribution from neutral bions and second term magnetic bions, in the second line we have the contributions from monopole-instantons. $V_{b,m}$ are constants defined by
\begin{equation*}
  V_{\text{b}} = \frac{27\Lambda^6}{8\pi m_W^3}\ln \left( \frac{m_W}{\Lambda} \right),\quad  
  V_{\text{m}}= \frac{9m \Lambda^3}{2\pi m_W} \ln \left( \frac{m_W}{\Lambda} \right)
\end{equation*}
and are fixed at large $N$. 
We will work in  the weak coupling limit
\[
  \frac{g^2 N}{8\pi^2} = [3\ln(m_W/\Lambda)]^{-1} \ll 1 \quad \text{since}\quad m_W \gg \Lambda.
\]

Following Ref.~\cite{Poppitz:2013} we now observe that we can neglect the perturbative contribution $V_{\rm p}$ to
the effective potential when $V_{\rm p} \ll V_{\rm m}$, which translates to the condition
\begin{align}
  m \ll \frac{\eta^2}{N}
  \label{eq:m_np_condition}
\end{align}
up to logarithmic corrections in $\eta$.  We will assume that this condition is
satisfied everywhere below, because otherwise the behavior of the holonomy would
be the same as in Yang-Mills theory at high temperatures, where all of the
eigenvalues of the holonomy are clumped together.

\begin{figure}
    \centering
    \includegraphics{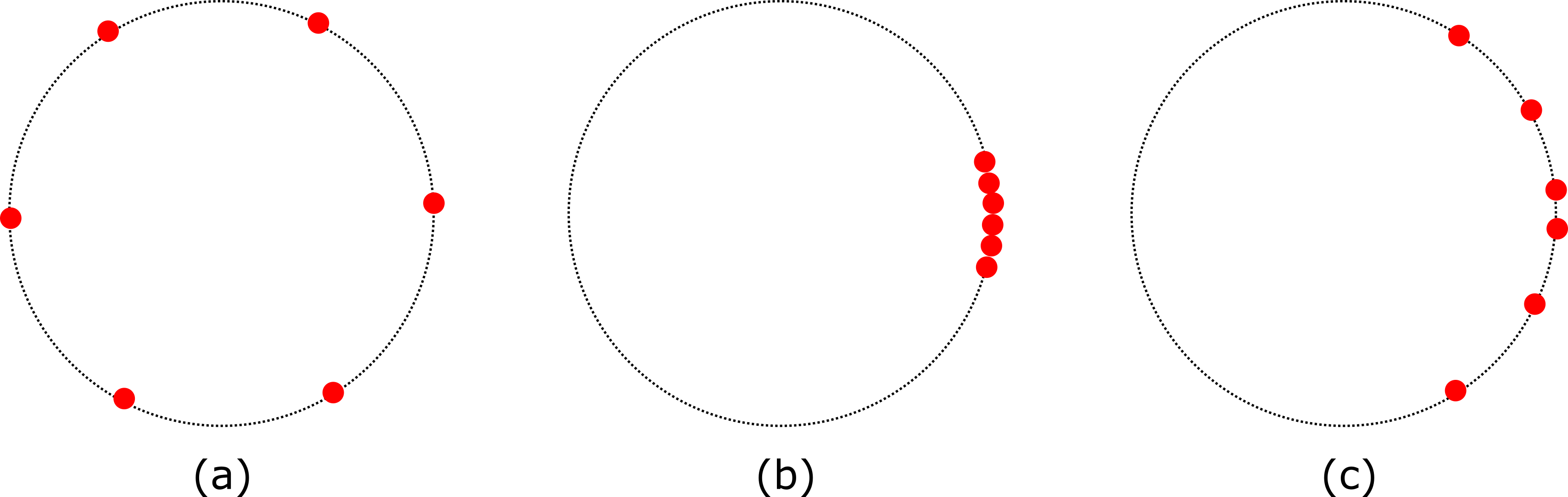}
    \caption{Possible distributions of the eigenvalues of the Polyakov loop. In panel (a) the eigenvalues are distributed homogeneously, rendering the trace of the Polyakov loop to be zero, hence the center symmetric is preserved. In panel (b) eigenvalues clumped at a single point, and the model undergoes complete center symmetry breaking. In panel (c) center symmetry is still broken but the eigenvalues have a non-equidistant distribution.}
    \label{fig:PolyakovLoopPhase}
\end{figure}

The effective potential Eq.~\eqref{eq:nppot} is much simplified when $m = 0$ since then the monopole contribution vanishes. We expect the ground state of $\sigma$ to be $0$, and when plugged as an ansatz into $V_{\rm np}$, we obtain 
\be
  V_{\text{np}} = V_{\rm b}\sum_{i=1}^N \left( e^{2a_i}-e^{a_i+a_{i+1}} \right)
\ee
where   ${a_i \equiv - \bm{\alpha_i}\cdot \bm{b} = b_{i+1}-b_i}$.  There exists a constraint $\sum_{i=1}^N a_i = 0$ as a result of identifying $b_{N+1} \equiv b_1$. Complete the squares
\begin{equation}
  V_{\text{np}} = V_{\rm b}\sum_{i=1}^N\left( \frac{1}{2}e^{2a_i} - e^{a_i}e^{a_{i+1}} +\frac{1}{2}e^{2a_{i+1}} \right)
  = \frac{1}{2}V_{\rm b} \sum_{i=1}^N\left( e^{a_i}-e^{a_{i+1}} \right)^2\ge 0.
  \label{eq:V0PD}
\end{equation}
we see that the minimum is obtained when $a_{i}=a_{i+1},\, i = 1,\cdots,N$. The constraint $\sum_{i=1}^N a_i = 0$ requires $a_i = b_{i+1}-b_i = 0$, which implies that $b_1=b_2=\dots =b_N$. Without loss of generality we can set $b_i = 0$ for all $i = 1,\cdots, N$, consequently the Polyakov loop has phases separated from each other by equal distances, as shown in Fig.~\ref{fig:PolyakovLoopPhase} panel~(a).  
This completes our review of the fact that 4d $\mathcal{N}=1$ SYM theory on a circle of size $L$ with periodic boundary conditions preserves center symmetry when $L$ is small.   The assumption that $L$ is small enough ensures that the calculation leading to Eq.~\eqref{eq:nppot} is reliable, and as we have stated before one must assume that $N L \Lambda \ll 1$.

If the fermion has a small but non-zero mass, the monopole contribution to the effective potential will be small but non-zero as well. Neglecting the $\sigma$ field for now, the monopole induced potential is proportional to $e^{\bm{\alpha}\cdot \bm{b}}$, similar to the bion contribution but with an extra minus sign.  
Thus there exists competition between the monopole and bion contribution where the strength of monopole contribution is controlled by the fermion mass \cite{Poppitz:2012sw,Poppitz:2013,Anber:2011de}.   If the fermion mass is large, the perturbative contribution to the holonomy effective potential dominates, and it is well-known that center symmetry is spontaneously broken.  But an imprint of this physics is visible even in the non-perturbative terms in Eq~\eqref{eq:nppot}, since the monopole-instanton terms look like $V_{\text{np}} \sim - m\sum e^{a_i}$, and if we can ignore the bion terms then the vacuum is obtained where $e^{a_i}$ are maximized, suggesting that center symmetry should be broken once the monopole terms become important enough.  All this implies that there must be a critical value of the fermion mass $m^{\text{cr}}$ where the system undergoes a center-symmetry-changing phase transition. 

In what follows we will briefly discuss this phase transition for small values of $N$.  This discussion has a close overlap with the results of  \cite{Poppitz:2012sw,Poppitz:2013}, but it is useful to go through it to set some notation.  We then turn a discussion of the phase transition at large $N$, with $\eta = NL\Lambda$ fixed and small. 

\subsection{Center symmetry breaking at small $N$}

\begin{figure}[htb]
  \centering
  \includegraphics[width=\linewidth]{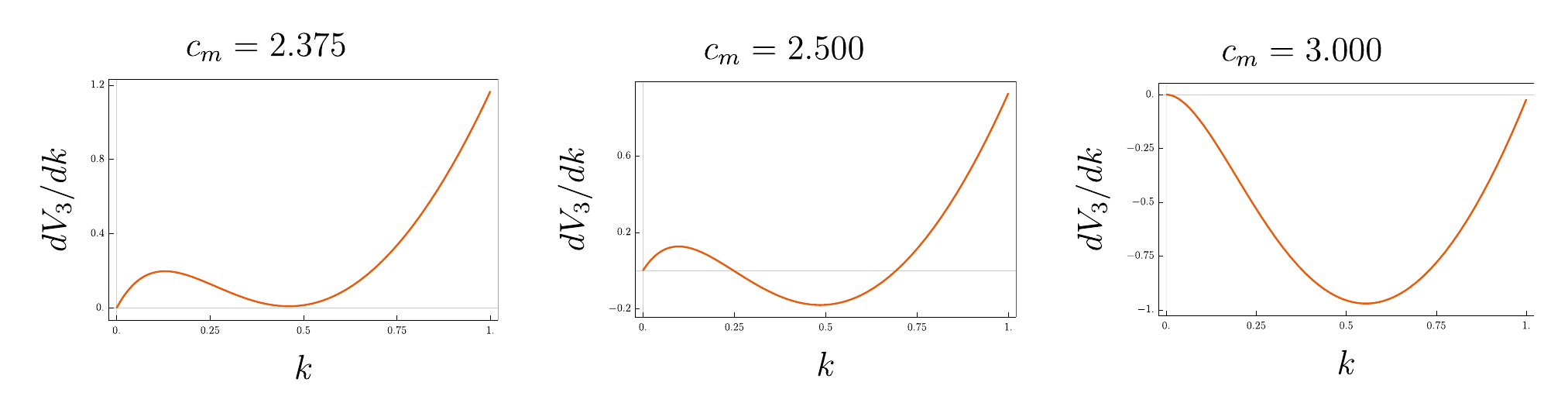}
  \caption{The value of $c_m$ increases from left to right. Left panel: two degenerate roots $k^*_{1}=k^*_{2}>0$. The center symmetric vacuum is stable globally. Middle panel: two distinct roots $0 < k^*_{1} < k^*_{2}0$. The center symmetry breaking phase transition occurs in this region. Right panel: There are two degenerate roots at $k=0 = k_1^*$.}
  \label{fig:dvs}
\end{figure}

The confinement phase transition takes place at the critical fermion mass $m^{\text{cr}}$.  For $N\le 4$ analytical expressions of $m^{\text{cr}}$ can be obtained, which is presented below. 
At $N=3$, only the differences of $b_{1,2,3}$ fields appear in the effective potential, thus we can add to them any constants so that $b_1 + b_2 + b_3 = 0$. Exploit the $\mathbb{Z}_2$ symmetry $i\to N-i$ we have $b_2-b_1 = b_3-b_2$ thus the parametrization $b_1 = -k,\, b_2 = 0,\, b_3 = k$ where $k \in \R$ could be adopted, the potential Eq.\eqref{eq:nppot} at $N=3$ reads
\be
  V_3 = V_{\rm b} \left[ e^{2k}+4e^{-4k}-2e^{-k}-c_m(2e^{k}+e^{-2k}) \right]
\ee
where
\be
  c_m \equiv \frac{V_{\text{m}}}{V_{\text{b}}} = \frac{4m m_W^2}{3\Lambda^3}
\ee
is a dimensionless parameter proportional to the fermion mass.  The minimum of  $V_3$ can be inferred from 
\be
  \partial V_3/\partial k = V_{\rm b} \left[ (2e^{2k}-4e^{-4k}+2e^{-k})-2c_m(e^{k}-e^{-2k}) \right] = 0
  \label{eq:dvdk},
\ee
{In the follow $k$ is always assume to take values in $\mathbb{R}$}. If $c_m$ is small enough, $\partial V_3/\partial k = 0$ has no solution for $k>0$, the only root is $k=0$, {meaning} that  $b_i =0$, and center symmetry is preserved. However if $c_m$ is large enough there exist two (possibly degenerate) roots at $k^*_{1,2}>0$. {The phase transition takes place when the effective potential at any of $k^*_{1,2}$ becomes smaller than that at $k=0$, however the center symmetric vacuum is still locally stable and the phase transition is of first order. At even larger $c_m \equiv c_m^*$ the center symmetric vacuum will {become} locally unstable. 
{The situation is summarized} in Fig.~\ref{fig:dvs}, {where} in the left panel there are two degenerate roots at $k>0$, in the middle panel there are two distinct roots, in the right panel there are two degenerate roots at $k=0$.}

Define $t\equiv e^k$ and reduced potential $\widetilde{V}_3 \equiv V_3/V_{\rm b}$, we have
\begin{align}
\notag
    \widetilde{V}_3(t) &= t^{-4} - 2 t^{-1} + t^2 - 2 c_m t - c_m t^{-2},\\
    \frac{\pd\widetilde{V}_3(t)}{\pd t} &= \frac{2}{t^5} \left(t^3-1\right) \left(t^3-c_m t^2 +2\right).
\end{align}
The stationary points of the effective potential are $t=1$, i.e. the trivial solution $b=0$, and the roots of $t^3-c_m t^2 +2=0$ with discriminant $\Delta = -108+8c_m^2$. The cubic equation has three roots if $c_m > \frac{3}{2^{1/3}}$, note that one of the three roots is negative thus irrelevant to our discussion. The other two roots corresponds to two stationary configurations of the $b$ field, to be more specific, the smaller positive root corresponds to the local maximum while the greater root is the global minimum. 
Denote the greatest root as $t_{\text{max}}$, we find
\be
t_{\text{max}} = \frac{1}{3}c_m +\frac{1}{3} \frac{c_m^3}{\left(-27+c_m^3+3\sqrt{81-6c_m^3}\right)^{1/3}} + \frac{1}{3}\left(-27+c_m^3+3\sqrt{81-6c_m^3}\right)^{1/3}.
\ee
The energy difference $\Delta V_3\equiv \widetilde{V}_3(0)-\widetilde{V}_3(t_{\text{max}})$ increases monotonically with $c_m$, from being negative to positive. The critical fermion mass at which phase transition occurs is given by $\Delta V_3=0$, which in turn implies
\be
c_m^{\text{cr}} = 6-6\sqrt{3} \sin\left(\frac{\pi}{9}\right),\quad N = 3.
\ee
This analytical result agrees with the numerical value $2.45$ given in Ref.~\cite{Poppitz:2013}.  
The analytical expressions for $b$ fields are rather cumbersome and we only present the numerical values,
\be
  b_1 = -0.631,\quad b_2 = 0, \quad b_3 = 0.631.
\ee

\begin{figure}[thb]
  \centering
  \begin{subfigure}[b]{0.45\linewidth}
    \centering
    \includegraphics[width=\linewidth]{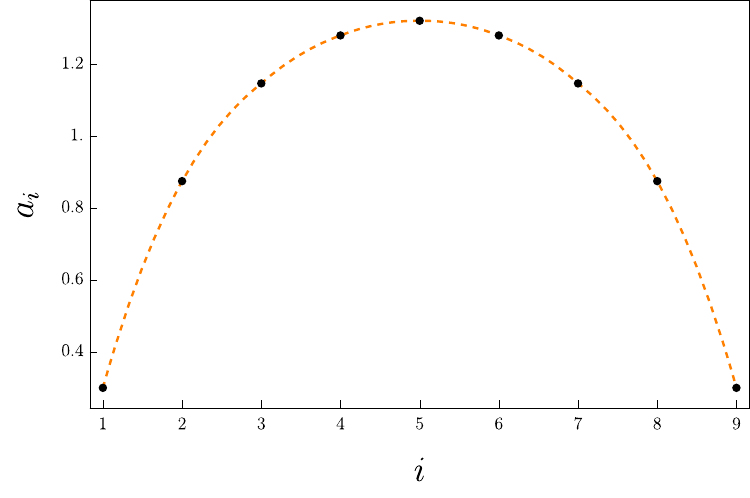}
    \caption{The value of $a_i$, $i = 1,\cdots,9$.}
  \end{subfigure}
  \hfill
  \begin{subfigure}[b]{0.45\linewidth}
    \centering
    \includegraphics[width=0.98\linewidth]{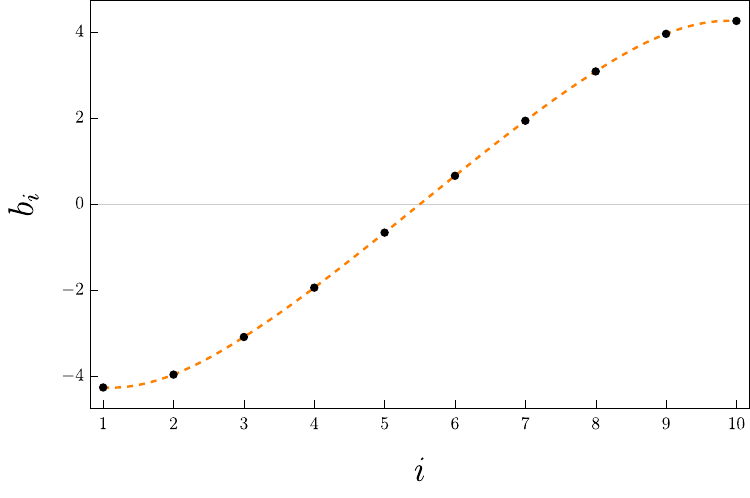}
    \caption{Normalized distribution of $b_i$ at $N = 10$.}
  \end{subfigure}
  \caption{The distribution of $a_i$ and $b_i$ at $N=10$, $c_m = 0.3$. The curves are extrapolations of numerical data points. $a_{10}$ is not shown in Panel (a) for it lies far below the x-axis. It shows that the emergent dimension adopts non-zero curvature when the center symmetry is broken.}
  \label{fig:aField10}
\end{figure}

The analytically expression of $c_m^{\text{cr}}$ for $N=4$ can be obtained but turns out to be quite cumbersome.  For $N>4$ in general no analytical expression of $c_m^{\text{cr}}$ can be obtained, thus numerical methods need to be adopted, with proper initial condition, such as a flat emergent dimension.  We can then vary $a_i$ to minimize the potential, with the constraint $\sum_{i=1}^N a_i = 0$.  Take $N=10$ for example, the critical fermion mass given by the numerical method is $c_m^{cr} \approx 0.24$,  
the normalized $b$ field distribution is shown in Fig.~\ref{fig:aField10} in the right panel. We see that the emergent space is not flat in the center symmetry broken phase. This is true in general for all $N>3$. Recall that $b$ is the deviation of the Polyakov loop phases from the center symmetric configuration, it means that the eigenvalues of the Polyakov loop now admit a non-trivial distribution.

\subsection{Center symmetry breaking at large \texorpdfstring{$N$}{N}}

For now the color space does not have the dimension of length. The fields, e.g., $b_i, \, i = 1, \dots, N$ are defined on a lattice with the lattice spacing defined to be $1/N$, at large $N$ the fields can be regarded as functions on a continuous color dimension.  In the large $N$ limit $\sum_{i=1}^N$ is replaced by $N \int_{0}^1 d x$ where $x\in [0,1]$, $a_i$ is replaced by $a(x) = \frac{1}{N}\frac{d b(x)}{dx}$ and $e^{a_i}-e^{a_{i+1}}$ is replaced by $\frac{1}{N}\partial_x e^{a(x)}$. 

Note that $b_{N+1} = b_1$ by construction, and in general the value of $b_N$ can not be identified to $b_1$, as a result $b(x)$ behaves as a step function at $x=1$ and $a(x) \propto \frac{d b(x)}{dx}$ behaves as a Dirac $\delta$ function at $x=1$. Indeed, this Dirac $\delta$ function is needed to satisfy the constraint $\int_0^1 a(x) = 0$ which is nothing but the continuous form of $\sum_i a_i =0$. This motivates us to write $a_i \to a(x)+C \delta(x-1)$, then $C$ can be chosen to be $(-\int a(x)dx)$.


After regrouping the $N$-summation into a continuous part and a Dirac $\delta$ function by rewriting the summation as $\sum_{i=1}^N = \sum_{i=1}^{N-2} + (\cdots)$,
in the continuous limit, $e^{a_i + a_{i+1}}$ can be substituted with $e^{2a(y)}$ and $ \sum_{i=1}^{N-2}$ becomes $N \int_0^1 dx$, the reduced potential reads
\begin{align}
\notag
    \widetilde{V}_{np} &= \frac{1}{2N} \int_{0}^1 dx\,\left[e^{2a(x)}(\partial_x a(x))^2- 2 N^2 c_m e^{a(x)}\right]\\
  &\;\;\;\, + \left[ e^{a_N}-\frac{1}{2}\left( e^{a(1)}+e^{a(0)} \right) \right]^2 + \frac{1}{4}\left( e^{a(1)}-e^{a(0)} \right)^2 -c_m e^{a(1)}-c_m e^{a_N},
  \label{eq:VContinuous}
\end{align}
where $a(0)$ is identified with $a_1$ and $a(1)$ is identified with $a_{N-1}$. Note the combined use of continuous and discrete labels. We can determine the distribution of eigenvalues by identifying the stationary points of the reduced potential. The continuous component of the potential, namely the expression given in the first line of Equation~\eqref{eq:VContinuous}, yields the Euler-Lagrange equation.
\be
  e^{a(x)}\left[\left(\frac{da}{dx}\right)^2+\frac{d^2 a}{dx^2}\right]+ \tilde{c} =0, \quad \tilde{c} \equiv {c_m}{N^2}.
  \label{eq:EoM}
\ee
Writing $a(x) = \ln(1+y(x))$ we find the polynomial solution for $y(x)$,
\be
  y(x) = -\frac{\tilde{c}}{2}x^2 + c_1 x + c_2
\ee
where $c_1, c_2$ are constants to be determined later. One of these two constants can be fixed by minimizing the integral in Eq.~\eqref{eq:VContinuous},
\be
  a(x) = \ln\left[1-\frac{\tilde{c}}{2}\left(x-\frac{1}{2}\right)^2 + c_3 \right].
  \label{eq:syma}
\ee

\begin{table}[tb]
  \centering
  \begin{tabular}{|p{15pt}|c|c|c|}
  \hline
  N & \textrm{Large N} &  Numerical \\
  \hline
  10 & 0.24  & 0.24   \\
  9  & 0.30 & 0.30 \\
  8  & 0.38 & 0.38 \\  
  7  & 0.49  & 0.50 \\
  6  & 0.67  & 0.68\\
   5  & 0.96 & 0.97 \\
   4  & 1.50 & 1.47  \\
  3  & 2.67 & 2.45 \\
  \hline
  \end{tabular}
  \caption{Comparison of analytical (large $N$ approximation) and numerical solutions for the critical value of the mass parameter $c_m = \frac{4m m_W^2 }{3\Lambda^3}$ in $\mathcal{N}=1$ SYM theory with a gluino mass $m$.  The large $N$ limit is taken such that $\tilde{m} = m N^2/\Lambda$ and $m_W/\Lambda = \frac{2\pi}{NL\Lambda}  = 1/\eta$ are held fixed as a function of $N$, and $\eta \ll 1$.}
  \label{table:cmComparison}
\end{table}

To fix $c_3$ the end-point terms need to be included. The constraint $\sum_i a_i =0$ implies that  $a_N = -N \int_0^1 dx a(x)\equiv -N S_a$ where $S_a$ is the signed area of the graph of $a(x)$. 
A direct calculation shows that
\be
 S_a = \ln \left(c_3-\frac{\tilde{c}}{8}+1\right)+\frac{4 \sqrt{2} \sqrt{c_3+1}}{\sqrt{\tilde{c}}}\tanh^{-1}\left(\frac{\sqrt{\tilde{c}}}{2 \sqrt{2} \sqrt{c_3+1}}\right)-2.
\ee
From Eq.~\eqref{eq:syma} we have $a(0)=a(1)=\ln(1+c_3-\tilde{c}/8)$ and the potential Eq.~\eqref{eq:VContinuous} reads
\be
 \widetilde{V}_{np}=\frac{\tilde{c}}{12 N}\left(\tilde{c}-12(c_3+1)\right)
 +\left( e^{-N S_a}-e^{a(0)}\right)^2 - c_m e^{a(0)}-c_m e^{-N S_a}.
\ee
In the center symmetry broken phase, the $a$ fields acquire non-zero vev thus $S_a >0$.  Neglect terms proportional to  $e^{-N S_a}$, we have
\begin{align}
c_3 
= -1+\frac{\tilde{c}}{8}.
\end{align}


\begin{figure}
    \centering
    \begin{subfigure}[t]{0.45\textwidth}
        \centering
        \includegraphics[width=0.9\linewidth]{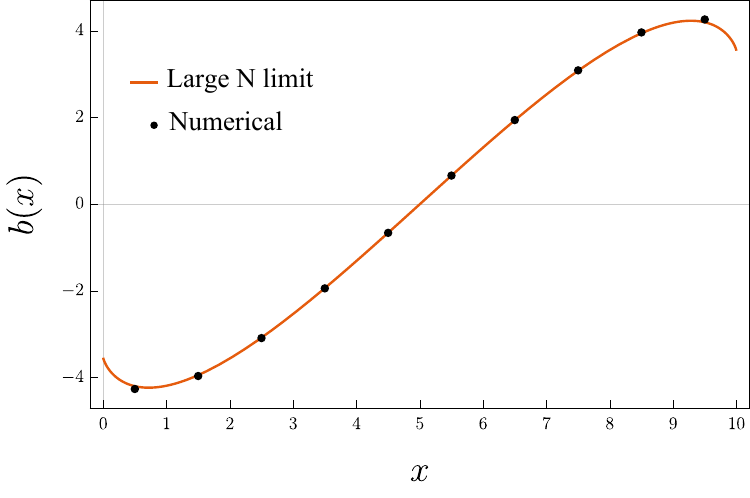} 
        \caption{$c_m$ scales as ${1/N^2}$, $N =10$, evaluated at $c_m = 0.3$. } \label{fig:bCompare1}
    \end{subfigure}
    \hfill
    \begin{subfigure}[t]{0.45\textwidth}
        \centering
        \includegraphics[width=0.9\linewidth]{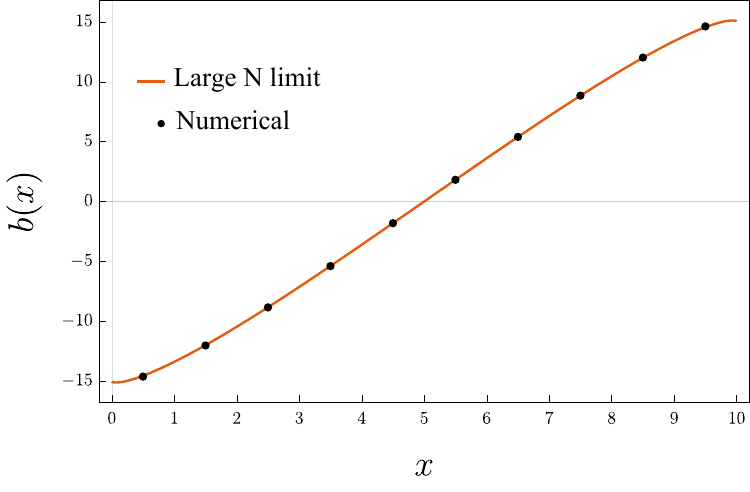} 
        \caption{$c_m$ scales as ${1/N}$, $N =10$, evaluated at $c_m = 3.0$. } \label{fig:bCompare2}
    \end{subfigure}
    \vspace{0.5cm}
    \begin{subfigure}[t]{0.45\textwidth}
        \centering
        \includegraphics[width=0.9\linewidth]{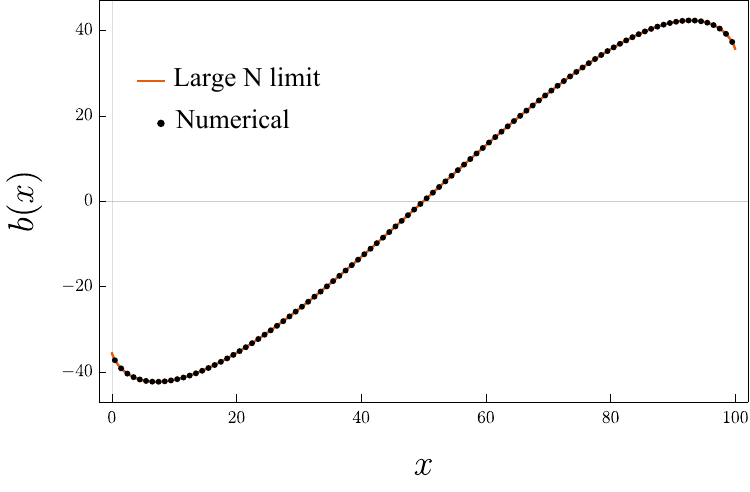} 
        \caption{$c_m$ scales as ${1/N^2}$, $N =100$, evaluated at $c_m = 0.003$.} \label{fig:bCompare3}
    \end{subfigure}
    \hfill
    \begin{subfigure}[t]{0.45\textwidth}
    \centering
    \includegraphics[width=0.9\linewidth]{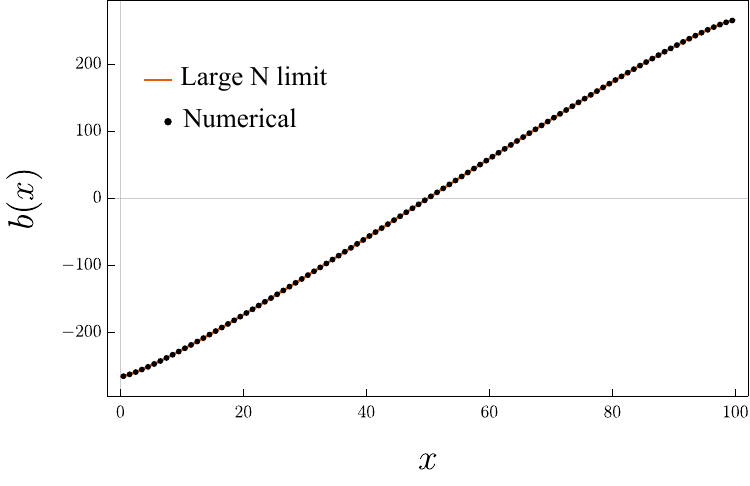} 
    \caption{$c_m$ scales as ${1/N}$, $N =100$, evaluated at $c_m = 0.3$.} \label{fig:bCompare4}
    \end{subfigure}
    \caption{In Panel~(a,b) we compare the numerical results of $b_i$ (dots) with the large-N analytical result (solid line) at $N=10$ with different scalings of the fermion mass. Note that for the numerical results, we have shifted the index from $\{i=1,\cdots, 10\}$ to $\{i=0.5,\cdots, 9.5\}$ so that it is odd with respect to $y=0.5$.
    In Panel~(a) the peculiar behavior of the curve at $x \lesssim 1$ and $x \gtrsim 1$ suggests the limitation of the continuum approximation method near the boundaries of the emergent space. In Panel (c,d) the comparison is shown for $N=100$. We find that as the fermion mass increases, in general the emergent dimension becomes more flat.}
    \label{fig:N10}
\end{figure}

Putting everything together, we have
\be
  a(x) = \ln\frac{\tilde{c}}{8}+\ln\left[1 -4\left(x-\frac{1}{2}\right)^2 \right].
  \label{eq:symaa}
\ee
Since  $a(x)=\frac{1}{N}\frac{d b(x)}{dx}$ we have
\begin{align}
  b(y) &= b(0) + N\left\{ y\ln\left[ \frac{\tilde{c}}{2}y(1-y)\right]-\ln(1-y)-2y  \right\}, \\
  b(1) &= b(0) + N\left( \ln\frac{\tilde{c}}{2}-2 \right).
  \label{eq:N10Con}
\end{align}
In Fig.~\ref{fig:N10} we compare the large-N analytical expression with the numerical results for $N=10$ and $100$ and $c_m \sim \order{1/N^2}, \order{1/N}$, while the other parameters are kept fixed. Recall that $c_m$ is required to be no larger than $\order{1/N}$ so the perturbative contribution to the effective potential is negligible. 
In Fig.~\ref{fig:N10} in order to make the discrete numerical result satisfy the same $\mathbb{Z}_2$ reflective symmetry as the large-N analytical distribution, i.e., $b(x)$ being an odd function with respect to the middle point of the emergent dimension, we have shifted the numerical results to the left by $0.5$. The figures show that at different orders of the fermions mass and different $N$ considered therein, the agreement is rather good. We find that as $c_m$ increases, in general the emergent dimension become more flat.

{Return} to the effective potential, in the large $N$ limit, in the center-breaking phase it is proportional to $m^2 N^3$ at the leading order,
\begin{align}
 V_{\textrm{broken, large N}}&=-\frac{\tilde{c}^2}{24 N}.
 \label{eq:pana}
\end{align}
In the center symmetric phase we have $a(x) \equiv 0$, the potential takes the value $V_{\textrm{unbroken, large N}} = -\tilde{c}/N$, this combined with Eq.~\eqref{eq:pana} implies that the critical value of $\tilde{c}$ at large $N$ is
\be
  \tilde{c}^{cr}= 24 \,, \;\; \textrm{large } N,
  \label{eq:crc}
\ee
which is independent of $N$. We compare the large $N$ result to the numerical solutions in Table~\ref{table:cmComparison} for $N =3, 4, \ldots, 10$ and found a rather good agreement.

\subsection{Effective action for the dual photon}
We now discuss the effective field theory describing fluctuations in the background of curved emergent dimension.
There are three fields to consider: the dual photons denoted by $\sigma$, the deviation $b(x^i,y)$ of the holonomy from the center-symmetric VEV and the fluctuations of the gluino fields.  At the supersymmetric point $m=0$, these three fields are related by supersymmetry, and so they have the same spectrum.  Away from the SUSY point, however, the dual photon field must be the lightest one.  The potential energies for the dual photons, holonomy fluctuations and gluino fluctuations are all zero at tree level.  But it is known that the potential energy of the dual photons cannot have any perturbative contributions, and so this potential is necessarily exponentially small compared to $m_W$ when the eigenvalues of the holonomy are not coincident.  There is no such restriction for effective potentials for the holonomy or gluino fluctuations, and for generic values of $m$ they will be heavier than the dual photons.

We take this as motivation to focus on the behavior of the dual photons in what follows.  We are considering a weakly coupled large $N$ limit, which means that e.g. self-interactions of $\sigma_i$ are suppressed by powers of our small parameters $1/N$ and $\eta$, as are their interactions with fluctuations of the holonomy and the gluinos.    
Before beginning the detailed analysis, we note that Eq.~\eqref{eq:nppot} implies that  dual photon potential contains terms that look like $f_i \cos[\alpha_i\cdot \sigma]$ and $g_i \cos((\alpha_i - \alpha_{i+1})\cdot \sigma)$.  
This means that at quadratic order in $\sigma_i$  we find terms of the form  $f_i \left( \sigma_{i+1}- \sigma_{i} \right)^2$ and $g_i \left(\sigma_{i+1}+\sigma_{i-1}- 2\sigma_{i} \right)^2$.  In the large $N$ limit the ``color'' index $i$ can be interpreted as a continuous coordinate, then for example $f_i [(\sigma_{i+1}- \sigma_{i})/a]^2$ in the continuous limit becomes $ f(y)(d\sigma/dy)^2$ with some scale $a$.  We will see below that the function $f$ is a constant when $\tilde{c} < \tilde{c}^{\rm cr} = 24$, but develops non-trivial dependence on $y$ when $\tilde{c} > \tilde{c}^{\rm cr}$, leading to an effective field theory for the dual photons that looks like a scalar field living in a curved 4d spacetime.

The effective potential at $\order{\sigma^2}$ takes the form
\be
  S \supset \int_{\R^3} dx^3 \sum_{i=1}^N \left( \frac{V_{\rm m}}{2}e^{a_i} (\sigma_{i+1}-\sigma_i)^2 + \frac{V_{\rm b}}{2}e^{a_i+a_{i+1}} (\sigma_{i+1}+\sigma_{i-1}-2\sigma_i)^2  \right),
  \label{eq:discrete}
\ee
in the large $N$ limit we have 
\be
  S \supset \int_{\R^3} dx^3 \int_0^1 dy  \left( \frac{V_{\rm b}}{2N^3}e^{2a(y)}\sigma''(y)^2
  + \frac{V_{\rm m}}{2N}e^{a(y)}\sigma'(y)^2 \right)
  \label{eq:actionbm}
\ee
where $\sigma'(y)\equiv d\sigma(y)/dy$. The bion contribution i.e., the term that contains $V_{\rm b}$ scales as $1/N^3$ while the instanton-monopole contribution containing $V_{\rm m}$ scales as $m/N$.  Recall that $m^{\text{cr}} \sim 1/N^2$, at the transition point both terms have the same large $N$ dependence.  
In terms of $\tilde{c}$ which is $N$-independent at large $N$, the instanton-monopole and bion contributions to the potential take the form 
\begin{align}
  S_m &=  \frac{V_{\rm b} \tilde{c}}{2N^3} \int_{\R^3} dx^3 \int_0^1 dy\,   g(y)\left(\frac{d\sigma(y)}{dy}\right)^2,
  \label{eq:twoSpatialDerivative}\\
  S_b &= \frac{V_{\rm b}}{2N^3}  \int_{\R^3} dx^3 \int_0^1 dy\, g^2(y)\left(\frac{d^2\sigma(y)}{dy^2}\right)^2
  \label{eq:fourSpatialDerivative}
\end{align}
where
\be
  g(y)= \begin{cases} 
  1 & \tilde{c} < 24  \\
  \frac{\tilde{c}}{8} \left( 1-4\left( y-\frac{1}{2} \right)^2 \right) & \tilde{c} > 24 
  \end{cases}
  \label{eq:aofy}
\ee
at the leading order, while the $\sigma$ kinetic term at large $N$ reads
\begin{equation}
S_\sigma =\int dx^3 \sum_i \lambda m_W (\partial_\mu\sigma_i)^2  \to \int dx^3 \int_0^1 dy N \, \lambda m_W (\partial_\mu\sigma(x^i,y))^2
\end{equation}
where $x^i$ are the coordinates on $\mathbb{R}^3$. 

The bion and instanton-monopole contribution can be estimated on the classical level without dimensional analysis. At the phase transition point $\tilde{c}^{\text{cr}} \approx 24$, $g(y)$ tends to zero near the boundary of the emergent space thus $g^2(y) \ll g(y)$, the bion potential with four spatial derivatives in Eq.~\eqref{eq:fourSpatialDerivative} is much smaller than the instanton-monopole potential with two spatial derivatives in Eq.~\eqref{eq:twoSpatialDerivative}. Near $y = 1/2$, due to the $\tilde{c}$ factor the instanton-monopole potential is still stronger than the bion potential by roughly a factor of $24$.

We can now rescale the coordinates and fields with dimensional parameters so that $y$ coordinate has dimension of length and the dual photon field is canonically normalized,
\be
  x\to x' = \xi x,\quad y \to y' = \zeta y,\quad \sigma \to  \sigma'={\Sigma}/{\mu}
\ee
where $\xi$ is dimensionless, $[\zeta] = [\mu] = \text{Length}^{-1}$. The action in terms of the rescaled fields reads
\begin{align}
\notag
  S = & \int dx'^3 \int_0^{1/\zeta} dy' \left\{
   {\frac{N \zeta \xi \lambda m_W}{\mu^2}}
   \left( \frac{\partial\Sigma}{\partial x'^\mu} \right)^2
   +g(\zeta y'){\frac{\xi^3 V_{\rm b} \tilde{c}}{2N^3\zeta\mu^2}}
   \left( \frac{\partial\Sigma}{\partial y'} \right)^2 \right.\\
   & \left. +g^2(\zeta y'){\frac{\xi^3 V_{\rm b}}{2N^3\zeta^3\mu^2}}
    \left( \frac{\partial^2\Sigma}{\partial y'^2} \right)^2
  \right\}.
  \label{eq:parametrization}
\end{align}
Setting $ \frac{N \zeta \xi \lambda m_W}{\mu^2}=1$ and $\frac{\xi^3 V_{\rm b}
\tilde{c}}{2N^3\zeta\mu^2} =1$ gives the desired normalization.  
Solving for $\mu$ and $\xi$ in terms of $\zeta$ using the condition $\xi^2 = \frac{2  \lambda  m_W \zeta^2 N^4}{\tilde{c} V_{\rm b} }$, and drop the primes we have
\be
  S = \int dx^3 \int_0^{1/\zeta} dy \left\{
   \left( \frac{\partial\Sigma}{\partial x^\mu} \right)^2
   +g(\zeta y)
   \left( \frac{\partial\Sigma}{\partial y} \right)^2
   +\frac{g^2(\zeta y)}{\Lambda_c^2}
    \left( \frac{\partial^2\Sigma}{\partial y^2} \right)^2
  \right\} 
  \label{eq:rescaledAction}
\ee
where $\Lambda_{c} = \sqrt{\tilde{c}}\zeta$.  The size of $y$
direction is $\tilde{L} \equiv 1/\zeta$.  
If we study physics on scales $\ell$
which satisfy $\Lambda_c^{-1} \ll \ell \le \tilde{L}$, the four-derivative term in
Eq.~\eqref{eq:rescaledAction} is technically irrelvant.  To be able to focus on
this regime, we need the hierarchy  $\Lambda_c \gg \zeta$, which in turn
requires that $\sqrt{\tilde{c}} \gg 1$.  Fortunately, at the center-symmetry
breaking transition we already have $\sqrt{\tilde{c}^\text{cr}} = \sqrt{24} \sim 5$, and
the necessary hierarchy becomes parametrically well-established as we further
increase $m$. Indeed, we can make $\tilde{c} \sim N$ without violating our
assumption about the scaling of $m$ in Eq.~\eqref{eq:m_np_condition}. We will
take these remarks as justification to neglect the four-derivative term in
Eq.~\eqref{eq:rescaledAction} in the analysis that follows, and will base our
analysis on 
\be
S = \int dx^3 \int_0^{\tilde{L}} dy \left\{
 \left( \frac{\partial\Sigma}{\partial x^\mu} \right)^2
 +g(y/\tilde{L}) 
 \left( \frac{\partial\Sigma}{\partial y} \right)^2 \right\}.
\ee
As a final comment in this section, we note that if we assume $\xi$ to be
independent of $N$, which amounts to studying distance scales that do not scale
with $N$, then we must scale $\zeta \sim 1/N^2$.  This means that the size of the
emergent dimension, $\tilde{L}$, scales as $N^2$.

\section{Effective Emergent Geometry at Large N}

The large $N$ action, dominated by the instanton-monopole contribution, can be recast into the action of Euclidean scalar field theory in the emergent curved background
\begin{equation}
    S = \int_M \dd x^3 \dd y \sqrt{|{g}|} g^{ab}\partial_a\phi \partial_b\phi
    \label{eqn:gencovscalaraction}
\end{equation}
where the base space is $M = \mathbb{R}^3\times (0,\tilde{L})$. This can be directly seen as follows. Consider the general form of the rescaled action, Eq.~\eqref{eq:rescaledAction} without the quartic term, 
\begin{equation}
    S = \int \dd x^3\int\dd y \left[(\partial_{\mu}\phi)^2 + f(y)(\partial_y \phi)^2 \right] = \int\dd x^3\int\dd y f(y)\left[\frac{1}{f(y)}{(\partial_\mu\phi)^2} + (\partial_y\phi)^2\right],
    \label{eqn:scalarftaction}
\end{equation}
the curved background emerges by constructing the inverse metric $g^{ab}$ by reading off the coefficients of the kinetic term together with demanding general covariance of the theory on $M$. We find the Euclidean metric 
\begin{equation}
    \dd s^2 = g_{ab}\dd x^a \dd x^b = f(y)\delta_{ij}\dd x^i\dd x^j + f(y)^{-1}\dd y^2,\hspace{.25in} f(y)^2 = \frac{\tilde{c}}{8}\left[1-4\left(\frac{y}{\tilde{L}}-\frac{1}{2}\right)^2\right] 
\end{equation}
with the Diff invariant measure $\dd \text{Vol}_g \sqrt{|{g}|} = \dd^3x\,\dd y f(y)$ identifies the actions Eq.~\eqref{eqn:scalarftaction} and Eq.~\eqref{eqn:gencovscalaraction}. To study the global structure and geodesic motion of particle excitations of the scalar field theory, we Wick rotate one of the flat Euclidean $x$-directions to the Lorentzian time, e.g. $x^3 \to \mathrm{i} t$. 
The Lorentzian metric reads
\be
\dd s^2 = f(y) (-\dd t^2 + \dd x_1^2 + \dd x_2^2) + f(y)^{-1} \dd y^2.
\label{eq:metric}
\ee
Compare with Eq.~\eqref{eqn:scalarftaction} we find that $f(y)^2 = g(y)$ where $g(y)$ is defined in Eq.~\eqref{eq:aofy}. 
The Riemann scalar curvature only dependents on $y$-coordinate,
\begin{equation}
    R = \frac{3 \tilde{c}^2 \left(4 y \tilde{L}+\tilde{L}^2-4 y^2\right)}{8 \sqrt{2} \tilde{L} \left(y \tilde{c} \left(\tilde{L}-y\right)\right)^{3/2}}
\end{equation}
and the Kretschmann scalar reads
\begin{equation}
  R_{\mu\nu\rho\sigma}R^{\mu\nu\rho\sigma}=\frac{3 \tilde{c} \left(\tilde{L}^2+4 y^2\right) \left(-8 y \tilde{L}+5 \tilde{L}^2+4 y^2\right)}{128 y^3 \tilde{L}^2 \left(\tilde{L}-y\right)^3}.
\end{equation}
We immediately n\textbf{}otice the singular behaviour of the scalar curvatures at $y=0, \tilde{L}$ i.e., the boundary of $y$ dimension, suggesting that these are not unphysical coordinates singularities, but rather singularities of the space-time. 
Recall that the emergent space is compactified by regarding the discrete color space as a lattice whose spacing is negligible in comparison with the total size, however, the color indices start with $1$ thus $y=0$ is not included in the continuous limit. The $N$-th index is dealt with separately from the continuous part, due to the $\delta$-function like behaviour, thus $y=\tilde{L}$ is not included in continuous $y$-space either. Hence, the singularities are excluded from the continuum and the emergent dimension has non-singular scalar curvature everywhere.

The metric, Eq.~\eqref{eq:metric}, turns out to be conformally flat. To see this one can either calculate the Weyl tensor directly to check that it is identically zero, or perform the coordinates transformation 
\be 
 {z}/{\tilde{L}} = -\frac{\pi}{4} + \arcsin \left( \sqrt{y/\tilde{L}} \right), \quad z/
\tilde{L}\in (-\pi/4, \pi/4),
\ee
the metric in terms of $z$ reads
\be
  ds^2 = f(z) (-dt^2 + dz^2 + dx_1^2 + dx_2^2)
  = f(z) (-dt^2 + dz^2 + dr^2 + r^2 d\theta^2)
  \label{eq:conMetric}
\ee
where $r, \theta$ are the radius and polar angle of the $(x_1,x_2)$ plane respectively, and 
\be
f(z) = \frac{\sqrt{\tilde{c}}}{2\sqrt{2}}\cos(2z/\tilde{L}).
\ee

\subsection{Geodesics of massive particles and light}

Should the metric adopt the Euclidean signature with coordinates $\{ x^1,x^2,x^3,z \}$, the conformal factor $f(z)$ can be interpreted as the optical index $n(\bm{x},z)$, the geodesic corresponds to the light path that minimized the optical path length with the optical index $n(\bm{x},z) = \sqrt{f(z)}$. For Lorentzian signature, the geodesic line can be found by solving geodesic equations.


For a massive particle, without making any assumptions there are four equations of motion 
\[
  \frac{d X^\lambda}{d\tau^2} + \Gamma^\lambda_{\mu\nu}\frac{dX^\mu}{d\tau} \frac{dX^\nu}{d\tau} = 0
\]
where $X$ are independent coordinates and $\tau$ the proper time. In our case the equations take the form 
\begin{align}
  &f(z)\frac{dt}{d\tau} = \epsilon,\label{eq:eqT}\\
  &\frac{d\theta}{d\tau} = \frac{l}{f(z) r^2},\label{eq:eqth}\\
  &f(z)\frac{d^2r}{d\tau^2} + f'(z)\frac{dr}{d\tau}\frac{dz}{d\tau} = 0, \\
  &\frac{2f(z)}{f'(z)} \frac{d^2 z}{d\tau^2} -\left( \frac{dr}{d\tau} \right)^2
  + \left( \frac{dt}{d\tau}\right)^2 + \left( \frac{dz}{d\tau} \right)^2 =0,
  \label{eq:eqZ}
\end{align}
where $\epsilon$ and $l$ are two integration constants associated with Killing vectors $\xi_E$ and $\xi_L$, which we will talk more about later. In addition to above equations of motion, there is the defining equation of proper time
\be
  \frac{1}{f(z)}  - \left( \frac{dt}{d\tau} \right)^2+ \left(\frac{dr}{d\tau} \right)^2 +  \left( \frac{dz}{d\tau} \right)^2 + r^2\left(\frac{d\theta}{d\tau}\right)^2  = 0.
  \label{eq:eqTau}
\ee
The above five equations combined together are more than enough to decide the trajectory of a massive particle.

The constant of integral $\epsilon$ is the conserved total energy $E$ per mass associated to the Killing vector $\xi^\mu_E = (1,0,0,0)$. To be more specific, $E$ and the angular momentum in the $x^1,x^2$ plane are
\begin{align}
E &= -m g_{0\mu}\,\xi^\mu_E = m f(z) \frac{dt}{d\tau}, \\
L &= m l
\end{align}
where $m$ is the mass of the particle, not to be confused with the mass of the adjoint fermion from previous sections. 
Similarly $l$ is the angular momentum per mass. Note that the geodesic is independent of the mass and only depends on the ratio $E/m, L/m$, as expected.

\begin{figure}[t]
  \centering
  \begin{subfigure}[b]{0.45\linewidth}
    \centering
    \includegraphics[width=\linewidth]{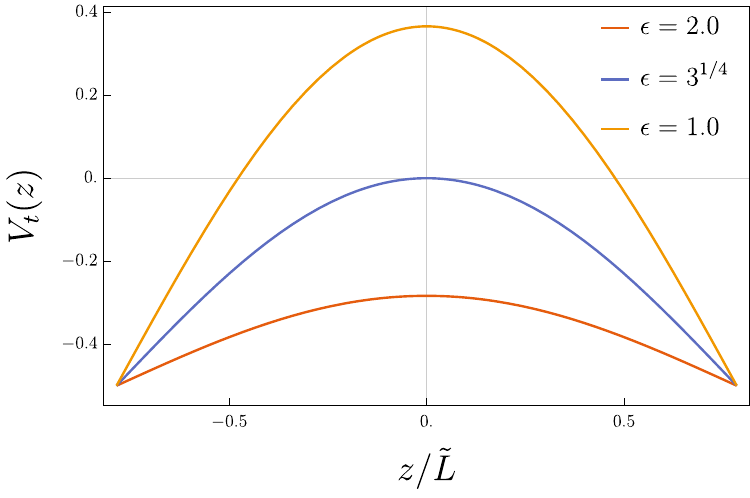}
    \caption{The effective potential in the emergent direction, $V_t(z) = {m^2}f(z)/{2E^2}-\frac{1}{2}$. }
  \end{subfigure}
  \hfill
  \begin{subfigure}[b]{0.45\linewidth}
    \centering
    \includegraphics[width=\linewidth]{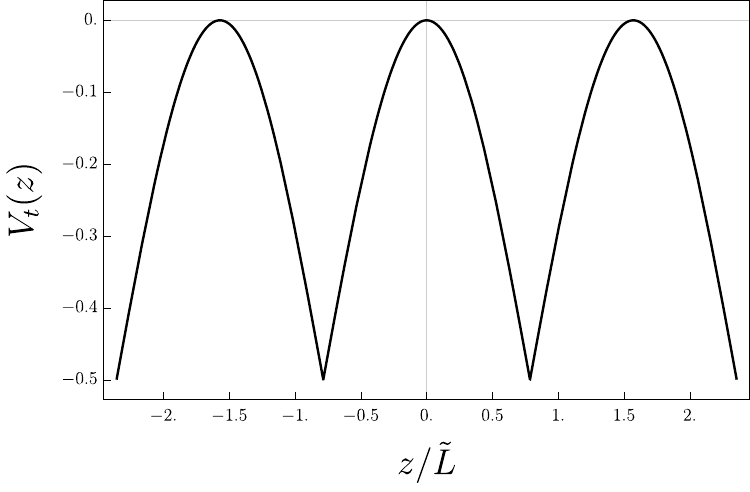}
    \caption{The compacted effective dimension at $\epsilon = 1.316$, $V_t(z=0) = 0$}
  \end{subfigure}
  \caption{The effective potential of massive particles in the large $N$ limit, defined by the geodesic equation $\frac{1}{2} \left( \frac{dz}{d t} \right)^2 + V_t(z) = 0$. In Panel (a) the {potential with maximum value above $V_t=0$ is forbidden since $V_t \leq 0$ is required}. Panel (b) shows the effective potential on an universal cover of the compactified emergent space {with identification $z/\tilde{L} \sim  z/\tilde{L}+\frac{\pi n}{2},\, n \in \mathbb{Z}$}. The effective potential is finite at the geometrically singular boundaries with cusps.}
  \label{fig:effPotZ}
\end{figure}

After some manipulation Eq.~\eqref{eq:eqTau} becomes
\be
  \frac{f(z)}{\epsilon^2}+\frac{l^2}{\epsilon^2 r^2}-1 +\left(\frac{dr}{dt}\right)^2+\left(\frac{dz}{dt}\right)^2 = 0,
\ee
focusing on the motion in the emergent dimension, we can further assume $r = \const$, from Eq.~(\ref{eq:eqT}, \ref{eq:eqZ}) we have
\be
  \frac{1}{2} \left( \frac{dz}{d t} \right)^2 + V_t(z) = 0,
  \quad V_t(z) = \frac{m^2}{2E^2}f(z)-\frac{1}{2},
  \label{eq:potZ}
\ee
which takes the same form as the equation of motion of a particle with unit mass in potential $V_t(z)$ with zero total energy. Note that Eq.~\eqref{eq:eqTau}, after rewriting $d/d\tau$ in terms of $d/dt$, gives the same effective potential up to an additive constant. 

From Fig.~\ref{fig:effPotZ} panel (a) we see that $\epsilon$ has a lower bound below which the kinetic energy become negative at certain $z$, which contradicts with Eq.\eqref{eq:potZ}. Since the maximum of potential energy is obtained at $z=0$, by requiring $V_t(z=0)<0$ we find the lower boundary $\epsilon \ge \epsilon_{\text{min}} = \left(\frac{\tilde{c}}{8}\right)^{\frac{1}{4}} \approx 1.316$ in the large $N$ limit. 

It is shown in Fig.~\ref{fig:effPotZ} panel (b) that, although the scalar tensor is singular at the boundary of $z$-space, the effective potential $V_t(z)$ {has well defined limits at $z=\pm \frac{\pi\tilde{L}}{4}$, making it possible for both the endpoints of the emergent dimension to be identified}, thus the full space is homeomorphic to $\mathbb{R}^3\times\s^1$. Note that the periodic effective potential is continuous but not smooth at the identification point. For instance, at one of the identification point $z/\tilde{L}=\pi/4$, the potential has the form
\begin{equation}
    V_t\left(\frac{\pi}{4}+\delta\right) = -\frac{1}{2} + \frac{\sqrt{\tilde{c}}|\delta|}{2\sqrt{2}\epsilon^2}.
\end{equation}

We note that the curved space-time can be detected by any particle that couples to the gauge holonomy. For instance, a heavy fermionic particle in the adjoint representation could serve as static test particle, provided it is heavy enough to effective decouple from the model.

\subsection{Causal structure}

\begin{figure}[t]
  \centering
  \begin{subfigure}[b]{0.45\linewidth}
    \centering
    \includegraphics[width=\linewidth]{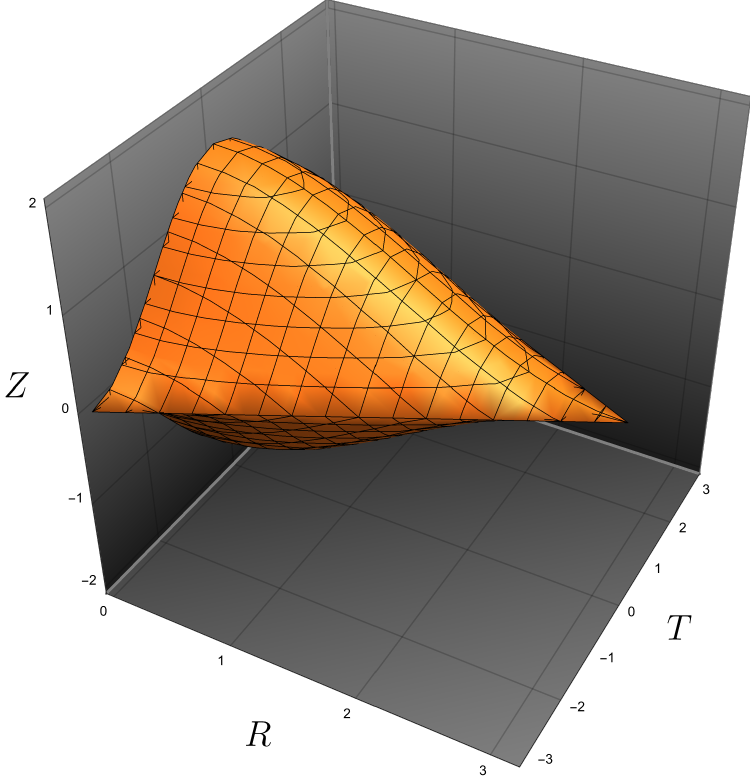}
    \caption{The 3D Penrose diagram. At $Z=0$ it coincide with the Penrose diagram of flat Minkowski space. The filled region is symmetric in $Z$ direction with respect to $z=0$ plane.}
  \end{subfigure}
  \hfill
  \begin{subfigure}[b]{0.45\linewidth}
    \centering
    \includegraphics[width=\linewidth]{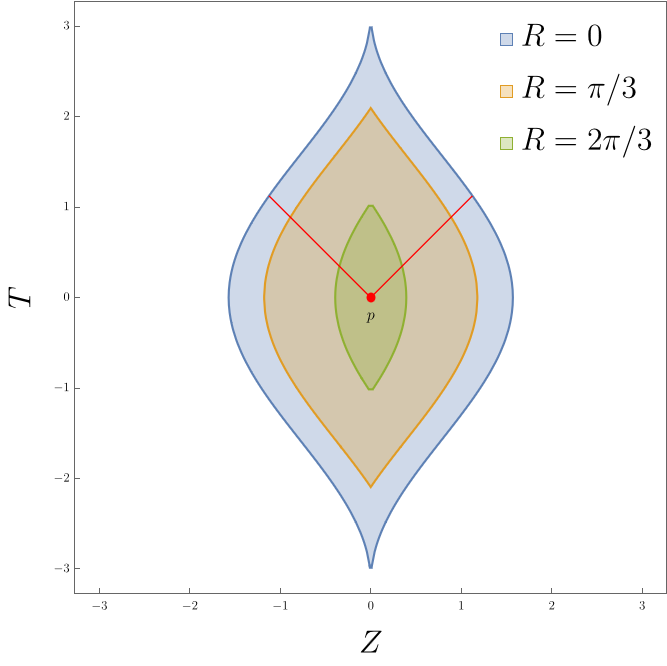}
    
    \caption{The profile of the Penrose diagram at various $R$. Given an event, e.g. $p$ at the center, the future light cone is given by the red straight lines.}
  \end{subfigure}
  \caption{Panel (a) shows the 3D Penrose diagram of $\mathbb{R}^3 \times \tilde{S}^1$ in coordinates $R,T,Z$, where $\tilde{S}^1$ is the emergent space, $R, T, Z$ are given by coordinate transformation $ t\pm r = \tilde{L} \tan\left( \frac{T \pm R}{2} \right)$ and $Z =  \left[\cos (T)+\cos (R)\right] {z}/{\tilde{L}}$.}
  \label{fig:Penrose}
\end{figure}

When the curvature is large, as is the case near the boundary in the emergent dimension, its effect on the causal structure of spacetime can no longer be neglected. In this section we discuss the causal structure illustrated by Penrose diagram~\cite{Penrose:1963}. A Penrose diagram should satisfy a) conformally flatness, so that the angles are preserved and the light rays always travel in $45^\circ$ and b) the infinities are compact, namely the entirety of spacetime is of finite graphic size, so that the causal structure anywhere in the spacetime can be read off from the diagram. This can be achieve by performing a specific conformal transformation from $(t,r,z)$ where $r = \sqrt{x_1^2 + x_2^2}$ to $R, T, Z$ given by 
\be
  t\pm r = \tilde{L} \tan\left( \frac{T \pm R}{2} \right), \quad Z =  \left[\cos (T)+\cos (R)\right] {z}/{\tilde{L}},
\ee
$\tilde{L}$ is needed for dimensional consideration. Note that the new coordinates $R, T, Z$ are dimensionless, in terms of which the metric adopts a concise form
\be
ds^2 = \frac{f(Z)\tilde{L}^2}{\alpha(R,T)^2}\left[-(dT)^2+(dR)^2+(dZ)^2\right], \quad
\alpha(R,T) = \cos (T)+\cos (R).
\ee
In the above expression the multiplicative factor before the bracket is not important as long as being semi-positive definite. Since a change of coordinates does not change the singularities, obviously the singularities will still appear at the boundary of now $Z$-dimension. This can be verified explicitly by evaluating the scalar curvature. 

The shape of the Penrose diagram is given by the range of $R, T, Z$,
\be
R\geq 0, \quad |T|\leq \pi -R, \quad |Z|<\frac{\pi}{4}\alpha (R,T)
\ee
where the first two relations yield the 2D Minkowski diagram.

The Penrose diagram is given in Fig.~\ref{fig:Penrose}. 
Since the spacetime we are studying is not spherically symmetric like Minkowski or Schwarzschild spacetime, a three dimensional diagram is needed to represent the entire space, as shown in panel (a) of Fig.~\ref{fig:Penrose}. 
At $Z=0$ we have recovered the Minkowski Penrose diagram (a solid triangle), although the timelike, lightlike and spacelike infinities are not labeled, they are not difficult to read off from the figure, e.g. the rightmost point is the spacelike infinity $i^0$.  
The spacetime is symmetric under $z\leftrightarrow -z$ and closes up as $Z, R$ or $T$ approaches the boundary. Making use of the property that light travel in $45^\circ$, panel (a) concisely shows the spacetime region accessible to observation. In Panel (b) the profile at fixed $R$ is shown, for an event in the spacetime e.g., $p$ at the center, the future light cone is given by the region upward within the red lines.

\section{Summary}

4d $\mathcal{N}=1$ super-Yang-Mills theory with gauge group $SU(N)$ deformed by $N_f=1$ massive adjoint fermion has so-called abelian large $N$ limit which is different form the usual 't Hooft large $N$ limit, in the sense that in the abelian limit $\eta = NL\Lambda$ is hold fixed at large $N$. The large distance effective model adopts center symmetry phase transition caused by the proliferation of topological defects, which we study in the paper using semi-classical methods. To account for the non-perturbative effects reliably, the mass of the adjoint fermion is required to be in a parametric window for the following reasons. 
If the fermion is too heavy, namely $m > 1/N$ then i) the perturbative contribution to the effective potential overwhelms the non-perturbative contribution and ii) the fermion-exchange induced interaction between instanton-monopoles takes a different form that the one used in the calculation. 
There exists a lower bound of the fermion mass because we want the higher derivative term in the effective Lagrangian in Eq.~\eqref{eq:rescaledAction} to be negligible. For $m < \order{1/N^2}$, the higher derivative term can no longer be neglected and the quantum corrections should be taken into consideration~\cite{Pisarski:2019} It is shown that the critical fermion mass $m^{\text{cr}}$ falls in the calculable window. 

In the center-symmetry broken phase, the curvature of ther emergent dimension is suppressed at heavy fermion mass, as shown in Fig.~\ref{fig:N10}. The size of the emergent dimension $\tilde{L}$ largely depends on the construction. For example, if we require that $\xi$ from Eq.~\eqref{eq:parametrization} is independent of $N$ then $\tilde{L}$ scales as $N^2$, which in terms of the size of the original $S_L^1$ reads $1/L^2$, which means large emergent size with assumptions made in the previous discussions.

The dual photon field $\sigma(\bm{x},y)$ owes its periodicity to the compatibility condition with the instanton-monopole configurations. At distance much larger than the size of the instanton-monopoles, the topological defects can be regarded as point-like particles created by $e^{i\sigma}$, which enters the effective Lagrangian as a degree of freedom \cite{Poppitz:2022}. 
The proliferation of instanton-monopoles and composite topological bions are responsible for the non-perturbative IR effective interaction.
In the center symmetry broken phase, the non-zero vacuum expectation value of $b(\bm{x},y)$ i.e., the deviation of Polyakov loop from the center symmetric vev, provides a non-trivial background to which the dual photon couples. As the result of center symmetry breaking, the emergent dimension is no longer homogeneous with the topology of a circle, instead non-zero curvature emerges and the topology breaks from $S^1$ to $\mathbb{R}$, with a $\mathbb{Z}_2$ reflection symmetry.

The curvature can be detected by a heavy test particle, such as adjoint fermions, for which the geodesic is given in the paper. The causal structure can be read off from the Penrose diagram shown in Fig.~\ref{fig:Penrose}. Note that at the boundary of the emergent dimension the scalar curvature diverges, which suggest that in the continuous limit, the boundary of the emergent space is not well defined. 
The analytical expression of the emergent curvature lays the ground for future studies, such as the consequences of the addition of adjoint fermions, or the study of the spectrum in the curved 4d effective model.

\acknowledgments
The authors express their sincere gratitude to Aleksey Cherman for bringing this project to their attention and for his insightful discussions. B.Z. would like to thank Pengming Zhang and Jarah Evslin for their many valuable discussions during the course of this work. B.Z. is supported by the Young Scientists Fund of the National Natural Science Foundation of China (Grant No.\ 12305079). A.\ D.\ was supported by the University of Minnesota Doctoral Dissertation Fellowship and by the National Science Foundation Graduate Research Fellowship under Grant No.\ 00039202.

\bibliographystyle{unsrt}
\bibliography{emergentDimension.bib}

\begin{thebibliography}{10}

\bibitem{tHooft:1973}
Gerard 't~Hooft.
\newblock {A Planar Diagram Theory for Strong Interactions}.
\newblock {\em Nucl. Phys. B}, 72:461, 1974.

\bibitem{Eguchi1982}
Tohru Eguchi and Hikaru Kawai.
\newblock Reduction of dynamical degrees of freedom in the large-iv gauge
  theory.
\newblock 48, 1982.

\bibitem{Bhanot1982}
Gyan Bhanot, Urs~M Heller, and Herbert Neuberger.
\newblock The quenched eguchi-kawai model, 1982.

\bibitem{Neuberger2020}
Herbert Neuberger.
\newblock Quenched eguchi-kawai model revisited.
\newblock {\em Physical Review D}, 102, 11 2020.

\bibitem{GonzalezArroyo1983}
A~Gonzalez-Arroyo and M~Okawa.
\newblock Twisted-eguchi-kawai model: A reduced model for large-n lattice gauge
  theory, 1983.

\bibitem{UnsalYaffe2010}
Mithat Ünsal and Laurence~G. Yaffe.
\newblock Large-n volume independence in conformal and confining gauge
  theories.
\newblock {\em Journal of High Energy Physics}, 2010, 2010.

\bibitem{Gross1982}
David J~Gross I and Yoshihisa Kitazawa.
\newblock A quenched momentum prescription for large-n theories.
\newblock {\em Nuclear Physics}, 206:440--472, 1982.

\bibitem{ArroyoOkawa2014}
Antonio González-Arroyo and Masanori Okawa.
\newblock Testing volume independence of su(n) pure gauge theories at large n.
\newblock {\em Journal of High Energy Physics}, 2014, 2014.

\bibitem{Arkani-Hamed:2001kyx}
Nima Arkani-Hamed, Andrew~G. Cohen, and Howard Georgi.
\newblock {(De)constructing dimensions}.
\newblock {\em Phys. Rev. Lett.}, 86:4757--4761, 2001.

\bibitem{Maldacena:1997re}
Juan~Martin Maldacena.
\newblock {The Large N limit of superconformal field theories and
  supergravity}.
\newblock {\em Adv. Theor. Math. Phys.}, 2:231--252, 1998.

\bibitem{Gubser:1998bc}
S.~S. Gubser, Igor~R. Klebanov, and Alexander~M. Polyakov.
\newblock {Gauge theory correlators from noncritical string theory}.
\newblock {\em Phys. Lett. B}, 428:105--114, 1998.

\bibitem{Witten:1998qj}
Edward Witten.
\newblock {Anti-de Sitter space and holography}.
\newblock {\em Adv. Theor. Math. Phys.}, 2:253--291, 1998.

\bibitem{DOUGLAS1995271}
Michael~R. Douglas and Stephen~H. Shenker.
\newblock Dynamics of su(n) supersymmetric gauge theory.
\newblock {\em Nuclear Physics B}, 447(2):271--296, 1995.

\bibitem{Unsal2009}
Mithat Ünsal.
\newblock Magnetic bion condensation: A new mechanism of confinement and mass
  gap in four dimensions.
\newblock {\em Physical Review D - Particles, Fields, Gravitation and
  Cosmology}, 80, 9 2009.

\bibitem{Cherman:2016jtu}
Aleksey Cherman and Erich Poppitz.
\newblock {Emergent dimensions and branes from large-$N$ confinement}.
\newblock {\em Phys. Rev. D}, 94(12):125008, 2016.

\bibitem{Aharony:1999ti}
Ofer Aharony, Steven~S. Gubser, Juan~Martin Maldacena, Hirosi Ooguri, and Yaron
  Oz.
\newblock {Large N field theories, string theory and gravity}.
\newblock {\em Phys. Rept.}, 323:183--386, 2000.

\bibitem{Polchinski:1992vg}
Joseph Polchinski.
\newblock {Strings and QCD?}
\newblock In {\em {International Symposium on Black holes, Membranes, Wormholes
  and Superstrings}}, 6 1992.

\bibitem{Girardello:1999bd}
L.~Girardello, M.~Petrini, M.~Porrati, and A.~Zaffaroni.
\newblock {The Supergravity dual of N=1 superYang-Mills theory}.
\newblock {\em Nucl. Phys. B}, 569:451--469, 2000.

\bibitem{Petrini:2018pjk}
Michela Petrini, Henning Samtleben, Stanislav Schmidt, and Kostas Skenderis.
\newblock {The 10d Uplift of the GPPZ Solution}.
\newblock {\em JHEP}, 07:026, 2018.

\bibitem{UnsalYaffe2008}
Mithat \"Unsal and Laurence~G. Yaffe.
\newblock Center-stabilized yang-mills theory: Confinement and large $n$ volume
  independence.
\newblock {\em Phys. Rev. D}, 78:065035, Sep 2008.

\bibitem{Poppitz:2012sw}
Erich Poppitz, Thomas Sch\"afer, and Mithat Unsal.
\newblock {Continuity, Deconfinement, and (Super) Yang-Mills Theory}.
\newblock {\em JHEP}, 10:115, 2012.

\bibitem{Anber:2014sda}
Mohamed~M. Anber and Tin Sulejmanpasic.
\newblock {The renormalon diagram in gauge theories on $
  {\mathrm{\mathbb{R}}}3\times {\mathbb{S}}1 $}.
\newblock {\em JHEP}, 01:139, 2015.

\bibitem{Poppitz:2012nz}
Erich Poppitz, Thomas Sch\"afer, and Mithat \"Unsal.
\newblock {Universal mechanism of (semi-classical) deconfinement and
  theta-dependence for all simple groups}.
\newblock {\em JHEP}, 03:087, 2013.

\bibitem{Anber:2015wha}
Mohamed~M. Anber and Erich Poppitz.
\newblock {On the global structure of deformed Yang-Mills theory and QCD(adj)
  on $ {\mathrm{\mathbb{R}}}^3\times {\mathbb{S}}^1 $}.
\newblock {\em JHEP}, 10:051, 2015.

\bibitem{Davies:2000nw}
N.~Michael Davies, Timothy~J. Hollowood, and Valentin~V. Khoze.
\newblock {Monopoles, affine algebras and the gluino condensate}.
\newblock {\em J. Math. Phys.}, 44:3640--3656, 2003.

\bibitem{Seiberg:1996nz}
Nathan Seiberg and Edward Witten.
\newblock {Gauge dynamics and compactification to three-dimensions}.
\newblock In {\em {Conference on the Mathematical Beauty of Physics (In Memory
  of C. Itzykson)}}, pages 333--366, 6 1996.

\bibitem{Bergner:2018unx}
Georg Bergner, Stefano Piemonte, and Mithat \"Unsal.
\newblock {Adiabatic continuity and confinement in supersymmetric Yang-Mills
  theory on the lattice}.
\newblock {\em JHEP}, 11:092, 2018.

\bibitem{Athenodorou:2020clr}
Andreas Athenodorou, Marco Cardinali, and Massimo D'Elia.
\newblock {Spectrum of trace deformed Yang-Mills theories}.
\newblock {\em Phys. Rev. D}, 104(7):074510, 2021.

\bibitem{Poppitz:2021cxe}
Erich Poppitz.
\newblock {Notes on Confinement on R3 \texttimes{} S1: From
  Yang\textendash{}Mills, Super-Yang\textendash{}Mills, and QCD (adj) to
  QCD(F)}.
\newblock {\em Symmetry}, 14(1):180, 2022.

\bibitem{Turner:2019wnh}
Carl Turner.
\newblock {Dualities in 2+1 Dimensions}.
\newblock {\em PoS}, Modave2018:001, 2019.

\bibitem{Poppitz:2022}
Erich Poppitz.
\newblock Notes on confinement on r3 × s1: From yang–mills,
  super-yang–mills, and qcd (adj) to qcd(f).
\newblock {\em Symmetry}, 14(1):180, Jan 2022.

\bibitem{Gross:1980br}
David~J. Gross, Robert~D. Pisarski, and Laurence~G. Yaffe.
\newblock Qcd and instantons at finite temperature.
\newblock {\em Rev. Mod. Phys.}, 53:43--80, Jan 1981.

\bibitem{Poppitz:2013}
Erich Poppitz, Thomas Sch\"{a}fer, and Mithat \"{U}nsal.
\newblock Universal mechanism of (semi-classical) deconfinement and
  $\theta$-dependence for all simple groups.
\newblock {\em Journal of High Energy Physics}, 2013(3), Mar 2013.

\bibitem{Anber:2011de}
Mohamed~M. Anber and Erich Poppitz.
\newblock {Microscopic Structure of Magnetic Bions}.
\newblock {\em JHEP}, 06:136, 2011.

\bibitem{Penrose:1963}
Roger Penrose.
\newblock Asymptotic properties of fields and space-times.
\newblock {\em Phys. Rev. Lett.}, 10:66--68, Jan 1963.

\bibitem{Pisarski:2019}
Robert Pisarski, V.~Skokov, and A.~Tsvelik.
\newblock A pedagogical introduction to the lifshitz regime.
\newblock {\em Universe}, 5:48, 01 2019.

\end{thebibliography}

\end{document}